\documentstyle[12pt,lnfprepCMS,epsfig,amsmath,here]{article}
\newcommand{\be}{\begin{equation}}
\newcommand{\ee}{\end{equation}}
\def\simge{\mathrel{%
      \rlap{\raise 0.511ex \hbox{$>$}}{\lower 0.511ex \hbox{$\sim$}}}}
\def\simle{\mathrel{
      \rlap{\raise 0.511ex \hbox{$<$}}{\lower 0.511ex \hbox{$\sim$}}}}
\newcommand{\Header}{
     \begin{tabular}{rl}
     \hspace{-.3cm}
     \includegraphics{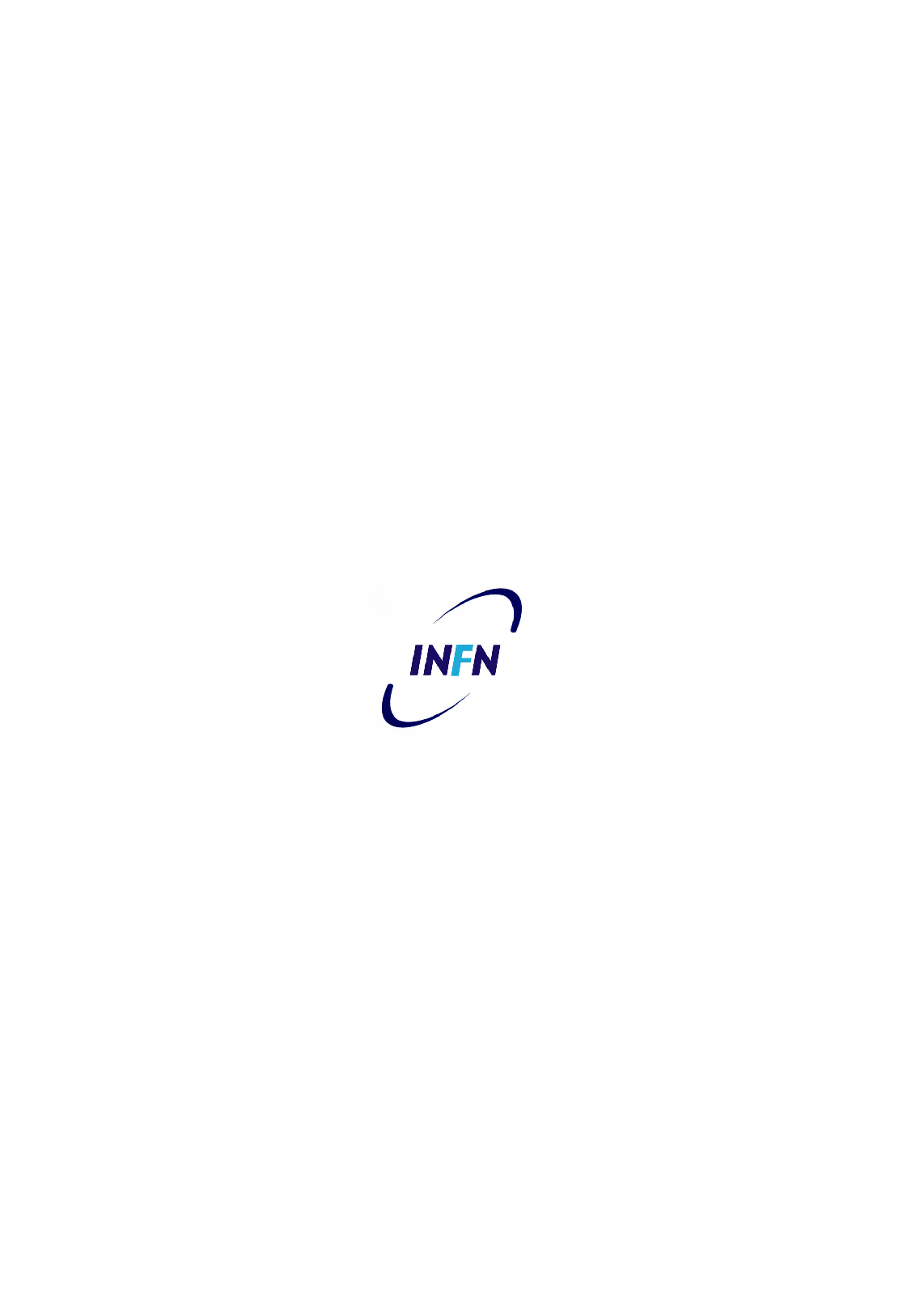} 
            &
       \renewcommand{\arraystretch}{0.5}
       \begin{tabular}{r}
         {\hspace{1cm}~\LARGE\sffamily LABORATORI~ NAZIONALI~ DI~ FRASCATI}\\
         \\
         {\Large\sffamily SIS-Pubblicazioni}\\
       \end{tabular}
       \renewcommand{\arraystretch}{1}
     \end{tabular}
  \vskip 0.5cm
  \begin{flushright}
  \renewcommand{\arraystretch}{0.5}
    \begin{tabular}{r}
      {\underline{LNF--08/29(P)}}\\    
      {\small November 29, 2008} \\      
      \\
    \end{tabular}
  \end{flushright}
  \renewcommand{\arraystretch}{1}
  \vskip 0.5 cm
  }

\begin{document}
\begin{titlepage}
\title{ 
  \Header
  {\large \bf THE CMS RPC GAS GAIN MONITORING SYSTEM:\\an Overview and Preliminary 
Results
}
}
\author{
L.~Benussi$^1$,
S.~Bianco$^1$,
S.~Colafranceschi$^1,2,3$,
D.~Colonna$^1$,
L.~Daniello$^1$, 
F.~L.~Fabbri$^1$,
M.~Giardoni$^1$
\\
B.~Ortenzi$^1$,
A.~Paolozzi,$^1,2$
L.~Passamonti$^1$,
D.~Pierluigi$^1$
\\
B.~Ponzio$^1$,
C.~Pucci$^1$,
A.~Russo$^1$,
G.~Roselli$^5$,
A.~Colaleo$^4$,
F.~Loddo$^4$,
M.~Maggi$^4$
\\
A.~Ranieri$^4$,
M.~Abbrescia$^4,5$,
G.~Iaselli$^4,5$,
B.~Marangelli$^4,5$,
S.~Natali$^4,5$
\\
S.~Nuzzo$^4,5$,   
G.Pugliese$^4,5$,
F.~Romano$^4,5$,
R.~Trentadue$^4,5$
\\
S.~Tupputi$^4,5$,
R.~Guida$^3$,
G.~Polese$^3,6$,
N.~Cavallo$^7$
A.~Cimmino$^7,8$
\\
D.~Lomidze$^8$,
P.~Noli$^7,8$
D.~Paolucci$^8$,
P.~Piccolo$^8$,
C.~Sciacca$^7,8$
P.~Baesso$^9$,
M.~Necchi$^9$
\\
D.~Pagano$^9$,
S.~P.~Ratti$^9$,
P.~Vitulo$^9$,
C.~Viviani$^9$\\
{\it ${}^{1)}$ INFN Laboratori Nazionali di Frascati, Via E. Fermi 40,
I-00044 Frascati, Italy.} \\
{\it ${}^{2)}$ Universit\`a degli Studi di Roma "La Sapienza",
Piazzale A. Moro.} \\
{\it ${}^{3)}$ CERN CH-1211 Gen\'eve 23 F-01631 Switzerland.} \\
{\it ${}^{4)}$ INFN Sezione di Bari, Via Amendola, 173I-70126 Bari,
Italy.} \\
{\it ${}^{5)}$ Dipartimento Interateneo di Fisica, Via Amendola,
173I-70126 Bari, Italy.} \\
{\it ${}^{6)}$  Lappeenranta University of Technology, P.O. Box 20
FI-538 1 Lappeenranta, Finland.} \\
{\it ${}^{7)}$ INFN Sezione di Napoli, Complesso Universitario di
Monte Sant'Angelo, edificio 6, 80126 Napoli, Italy.} \\
{\it ${}^{8)}$ Universit\`a di Napoli Federico II, Complesso
Universitario di Monte Sant'Angelo, edificio 6, 80126 Napoli,
Italy.}\\
{\it ${}^{9)}$ INFN Sezione di Pavia, Via Bassi 6, 27100 Pavia, Italy and Universit\`a degli studi di Pavia, Via Bassi 6, 27100 
Pavia, Italy.} 
}
\maketitle
\baselineskip=14pt

\begin{abstract}
The status of the CMS RPC Gas Gain Monitoring (GGM) system  developed 
at the Frascati Laboratory of INFN (Istituto Nazionale di Fisica Nucleare) is 
reported on. The GGM system is a cosmic ray telescope based on small RPC 
detectors operated with the same gas mixture used by the CMS RPC system. The GGM 
gain and efficiency are continuously monitored on-line, thus  providing a fast 
and accurate determination of any shift in working point conditions. The 
construction details and the first result of GGM commissioning are described.\\ 
\end{abstract}

\vspace*{\stretch{2}}
\begin{flushleft}
\end{flushleft}
\begin{center}

{\it{Presented by L.~Benussi at the RPC07 - February 2008, Mumbai, India}}
\end{center}
\end{titlepage}
\pagestyle{plain}
\setcounter{page}1
\baselineskip=17pt

\section{Introduction}
Design parameters, construction, prototyping and preliminary commissioning 
results of the CMS RPC Gas Gain Monitoring (GGM) system are presented 
\cite{Abbrescia:2007mu}.
The Resistive Plate Counter (RPC) 
system  is part of the muons detector of the Compact Solenoid Spectrometer (CMS) 
experiment\cite{cms} at the Large Hadron Collider (LHC) collider in CERN 
(Geneva, Switzerland), with the primary task of providing first level trigger 
and synchronization.
 The CMS 
RPCs are bakelite-based double-gap RPC with strip readout (for construction 
details see 
\cite{rpc_build} and reference therein) operated with 96.2\% C$_2$H$_2$F$_4$ - 
3.5\%
Iso-C$_4$H$_{10}$ - 0.3\% SF$_6$ gas mixture humidified at about 40\%. The large 
volume of the
whole CMS RPC system and the cost of gas used make mandatory the operation of 
RPC in a
closed-loop gas system (for a complete description see \cite{gassystem}), in
which the gas fluxing the gaps is reused after being purified by a set of 
filters\cite{purifiers}.

The operation of the CMS RPC system is strictly correlated to the ratio between 
the gas mixture components and to the presence of pollution due to contaminants 
that can be  be produced inside the gaps during discharges (i.e. HF produced by 
SF$_6$ or C$_2$H$_2$F$_4$
molecular break-up and further fluorine recombination), accumulated in the 
closed-loop or by
pollution that can be present in the gas piping system (tubes, valves, filters, 
bubblers, etc.)
and flushed into the gaps by the gas flow. To monitor the presence of these 
contaminants as
well as the gas mixture stability, is therefore mandatory to avoid RPC damage 
and to ensure their correct functionality. 

A monitoring system of the RPC working point due to changes of gas composition 
and pollution must provide a faster and sensitive response than the CMS RPC 
system itself in order
to avoid irreversible damage of the whole system. Such a Gas Gain Monitoring 
system  monitors
efficiency and signal charge continuously by means of a small sized cosmic ray 
telescope based
on RPC detectors. In the following will be briefly described the final setup of 
the GGM system, its construction details and
the first results obtained during  its commissioning done at the ISR area 
(CERN). 

\section{The Gas Gain Monitoring System}
The GGM system  is composed by the same type of RPC used in the CMS detector but 
of smaller size (2mm Bakelite gaps, 50$\times$50 cm$^2$).
Twelve gaps are arranged in a stack located in the CMS gas
area (SGX5 building) in the surface, close to CMS assembly hall (LHC-P5). The 
choice to install the
system in the surface instead of underground allows one to profit from maximum 
cosmic muon rates. In order to ensure a fast response to working point shifts 
with a precision
of 10\%, $10^3$ events are are required, corresponding to 
about 30 minutes exposure time on surface, to be compared with a 100-fold lesser 
rate underground. The trigger will be
provided by four out of twelve gaps of the stack, while the remaining eight gaps 
will be used to monitor the working point stability.  

The eight gaps are arranged in three sub-system: one sub-system (two gaps) will 
be fluxed with the fresh CMS mixture and its output sent to vent. The second 
sub-system (three gaps) will be fluxed with CMS gas coming from the closed-loop 
gas system and extracted before the gas purifiers, while the third sub-system 
(three gaps) are operated with CMS gas extracted from the closed-loop extracted 
after the gas filters.  
The basic idea is to compare the operation of the three sub-system and, if some 
changes are observed, to send a warning to the experiment. In this way, the gas 
going to and coming from the CMS RPC detector is monitored by using the two gaps 
fluxed with the fresh mixture as reference gaps. This setup will ensure that 
pressure, temperature and humidity changes affecting the gaps behavior do cancel 
out by comparing the response of the three sub-system operating in the same 
ambient condition. 

The monitoring is performed by measuring the charge distributions of each 
chamber. The eight gaps will be operated at different high voltages, fixed for 
each chamber, in order to monitor the total range of operating modes of the 
gaps. The operation mode of the RPC changes as a function of the voltage 
applied. A fraction of the eight gaps will work in pure avalanche mode, while 
the remaining will be operated in avalanche+streamer mode. Comparison of  signal 
charge distributions and the ratio of the avalanche to streamer components of 
the ADC provides a monitoring of the stability of working point for changes due 
to gas mixture variations.
\section{Construction and commissioning}
\begin{center}
  \begin{figure}[H]
    \resizebox{0.5\textwidth}{!}{\includegraphics{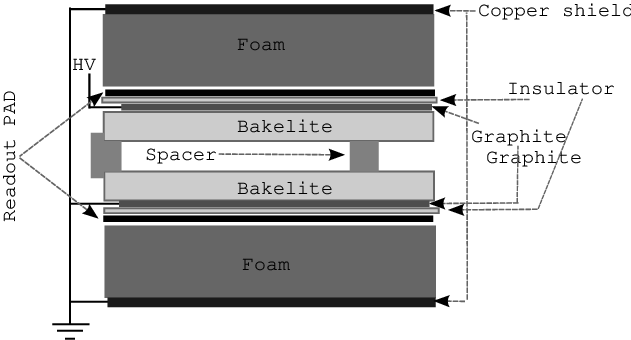}}
    \caption{A schematic layout of a GGM chamber}
    \label{fig1}
  \end{figure}
\end{center}

\begin{center}
  \begin{figure}[H]
    \resizebox{0.4\textwidth}{!}{\includegraphics{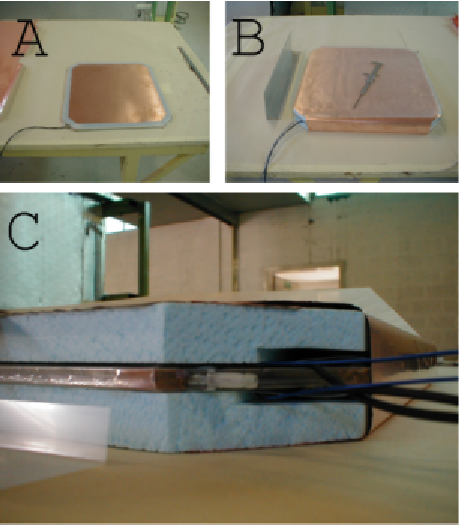}}
    \caption{Pictures of a GGM gap and chamber. A) a bare gap with the HV and 
signal cables. B)
      a completed chamber. The gap is sandwiched between two foam panels and 
fully covered with
      a copper shield. C) a section of a chamber with the two foam panels 
visible.}
    \label{fig2}
  \end{figure}
\end{center}

\begin{center}
  \begin{figure}[H]
    \resizebox{0.37\textwidth}{!}{\includegraphics{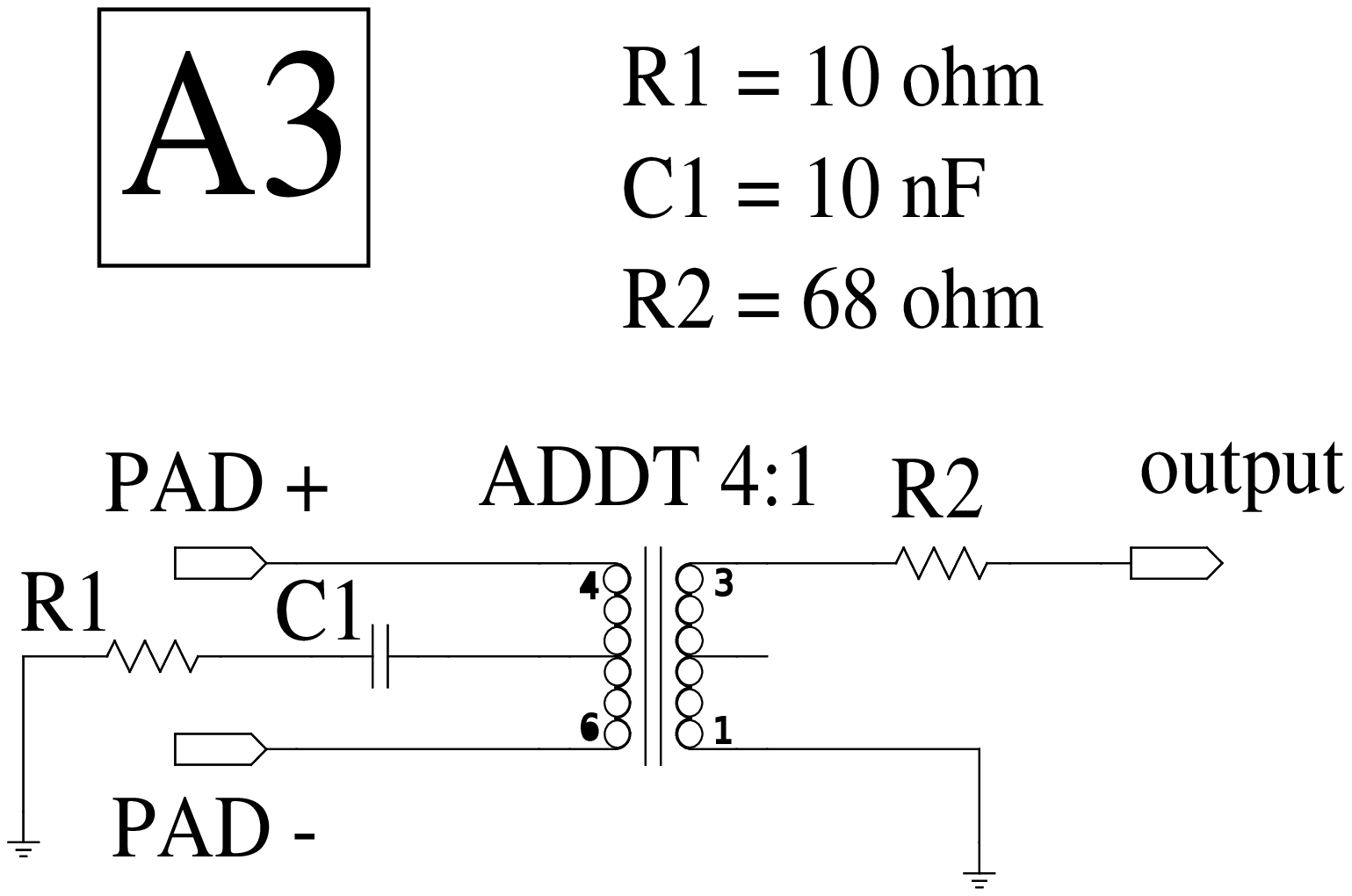}}
    \caption{The electric scheme of the read-out circuit providing the algebraic 
sum of the two      pad signal (PAD + and PAD -).}
    \label{fig3}
  \end{figure}
\end{center}

\begin{center}
  \begin{figure}[H]
    \resizebox{0.34\textwidth}{!}{\includegraphics{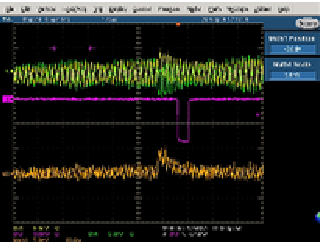}}
    \caption{An oscilloscope screen-shot of the two pad signals
      (upper traces) which are effected by a coherent noise and are barely 
visible on the screen. In
      the lower trace the coherent noise is highly reduced by A3 circuit. The 
vertical scale is the same for both cases 5 mV/div.}
    \label{fig4}
  \end{figure}
\end{center}

\begin{center}
  \begin{figure}[H]
    \resizebox{0.34\textwidth}{!}{\includegraphics{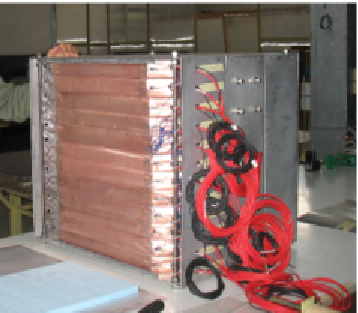}}
    \caption{A picture of the GGM system ready to be shipped to CERN. The stack 
is enclosed into
      an aluminum box for further shielding.}
    \label{fig5}
  \end{figure}
\end{center}

Each chamber of the GGM system consists of a single gap with double sided pad 
read-out: two copper pads are glued on the two opposite external side of the 
gap. Fig.\ref{fig1} shows a sketch of a chamber whose photos are shown in 
fig.\ref{fig2}; the two foam planes are used to
reduce the capacity coupling between the pad and the copper shields. The signal 
is read-out by a transformer based circuit A3 (Fig.\ref{fig3}). The 
circuit

allows to algebraically subtract the two signal, which have opposite polarities, 
and to obtain an output signal with subtraction of the coherent noise, with an 
improvement by about a factor 4 of the signal to noise ratio. 
Fig.\ref{fig4} shows the typical operation mode of the GGM double-pad readout 
with positive and negative pads pulses, and the output pulse from circuit A3.
The output signals from circuit A3 are sent to a CAEN V965 ADC \cite{caen} for 
charge analysis. 
The GGM has been tested with cosmic rays at LNF and then shipped to CERN for the 
final commissioning (fig.\ref{fig5} show the final stack assembly).

\begin{center}
  \begin{figure}[H]
   \resizebox{0.44\textwidth}{!}{\includegraphics{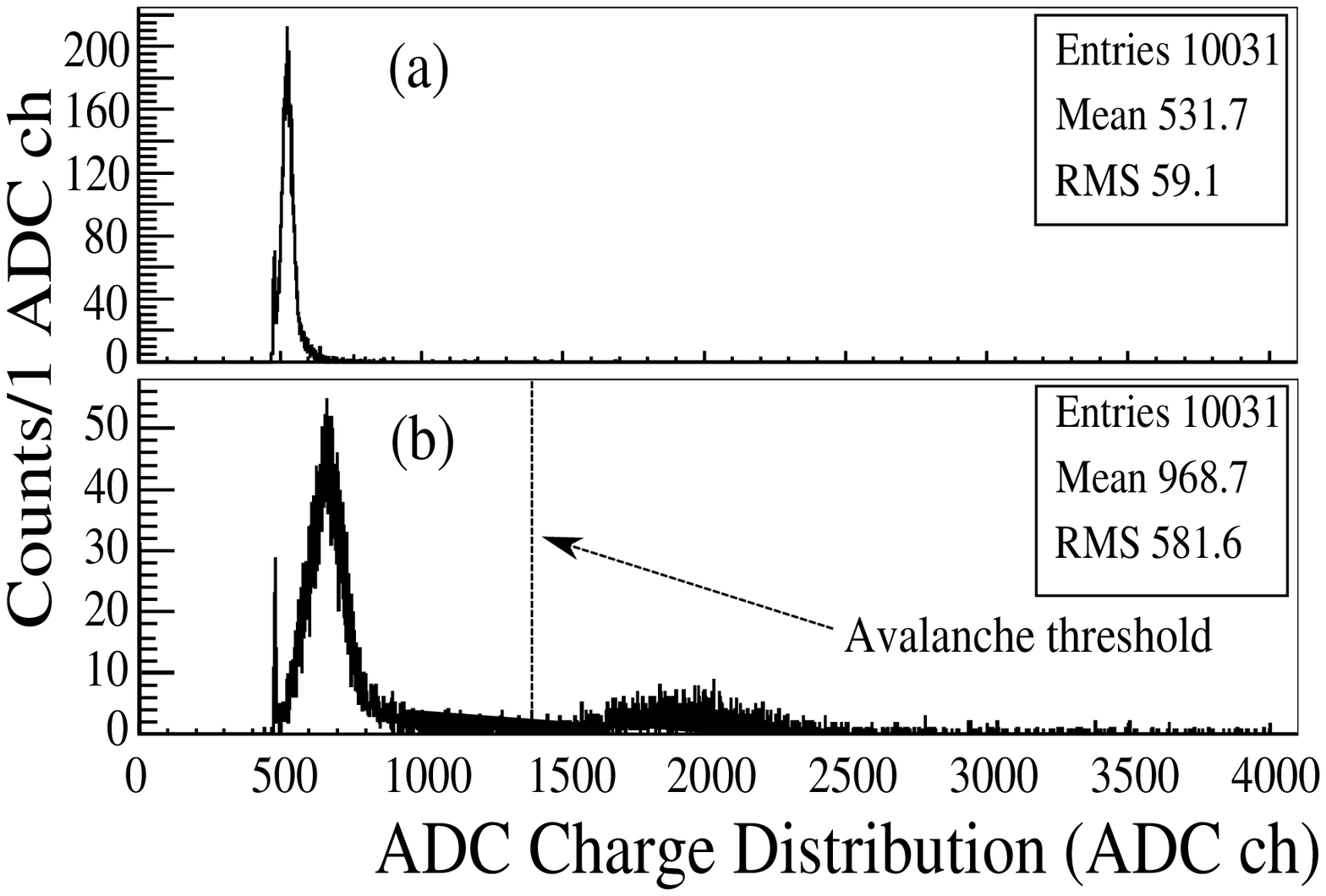}}
    \caption{Typical ADC charge distributions of one GGM chamber at two 
operating voltages. Distribution (a) correspond to HV$_{eff}$ = 9.9kV while 
distribution (b) to
      HV$_{eff}$=10.7kV. In (b) is clearly visible the streamer peak around 1900 
ADC channels. 
      The events on the left of the vertical line (1450 ADC channels in this 
case) are assumed to be pure avalanche events.}   
    \label{fig6}
  \end{figure}
\end{center}

A typical ADC distribution of a GGM gap is shown in fig.\ref{fig6} for two 
different effective
operating voltage,  defined as the high voltage set on the HV power supply 
corrected for the local
atmospheric pressure and temperature. Fig.\ref{fig6} a) corresponding to 
HV$_{eff}$=9.9kV shows a clean avalanche peak well separated from the pedestal. 
Fig.\ref{fig6} b) shows the charge distribution
at HV$_{eff}$=10.7kV with two signal regions corresponding to the avalanche and 
to avalanche+streamer mode.

\begin{center}
  \begin{figure}[H]
    \resizebox{0.44\textwidth}{!}{\includegraphics{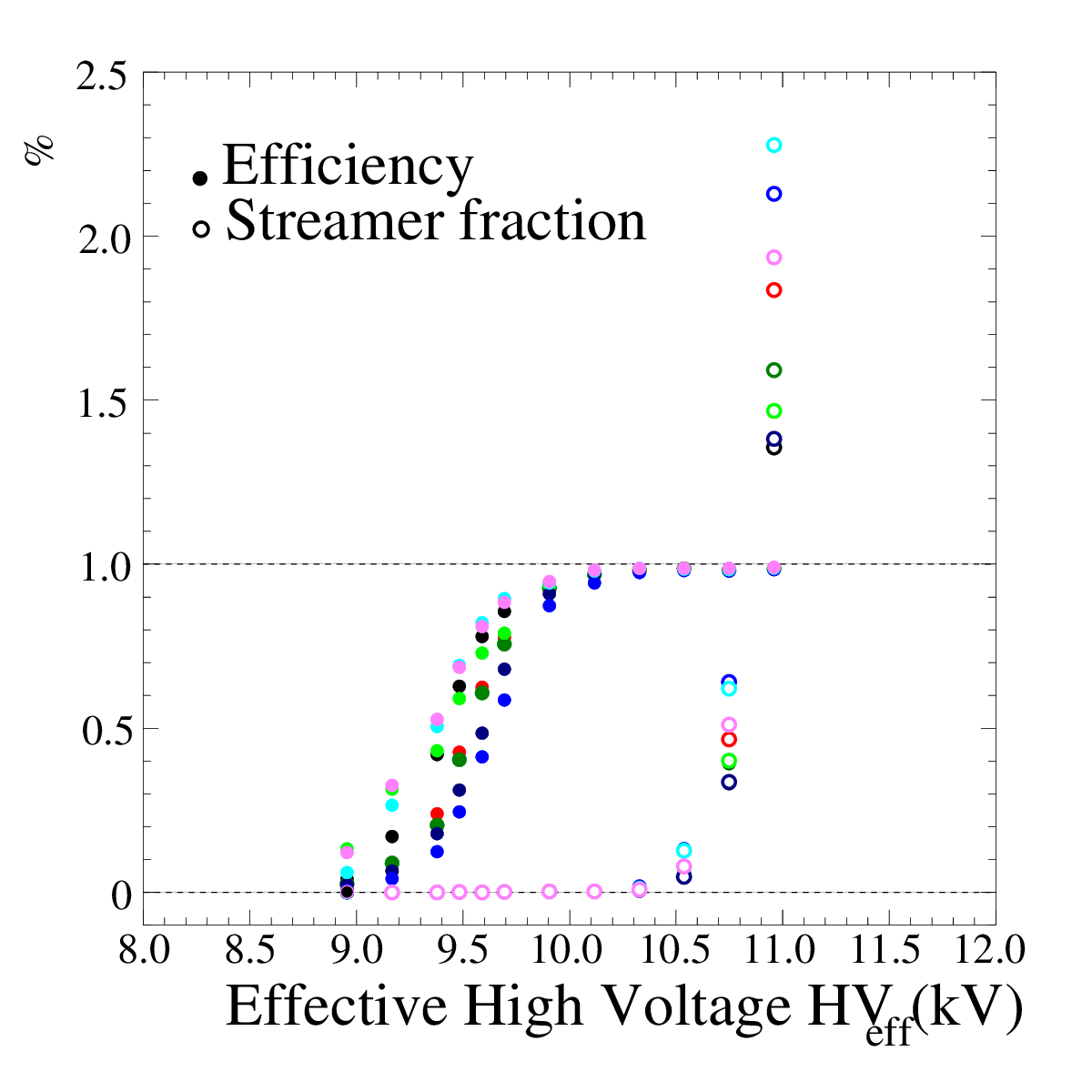}}
    \caption{Efficiency plot (full dots) of GGM chambers as a function of 
HV$_{eff}$. The efficiency is
      defined as the ratio between the number of ADC entries above 
3$\sigma_{ped}$ and the number
      of acquired triggers. Open dot plots correspond to the streamer fraction 
of the chamber
      signal as a function of HV$_{eff}$.}
    \label{fig7}
  \end{figure}
\end{center}
Fig.\ref{fig7} shows the GGMS single gap efficiency (full dots), and the ratio 
between the
avalanche and the streamer component (open circles), as a function of
the effective high voltage. Each point corresponds to a total of 10000 entries 
in the full ADC spectrum. The efficiency is defined as the ratio between the 
number of triggers divided by the
number of events above 3$\sigma_{ped}$ over ADC pedestal, where $\sigma_{ped}$
is the pedestal width. 
The avalanche to streamer ratio is defined
by counting the number of entries in the avalanche (below the ADC threshold 
(fig.\ref{fig6} b)
and above the pedestal region) and dividing it by the number of streamer events above the avalanche threshold. Both 
efficiency and avalanche plateau are in good agreement with previous results \cite{Abbrescia:2005yr}.

\begin{center}
  \begin{figure}[H]
    \resizebox{0.44\textwidth}{!}{\includegraphics{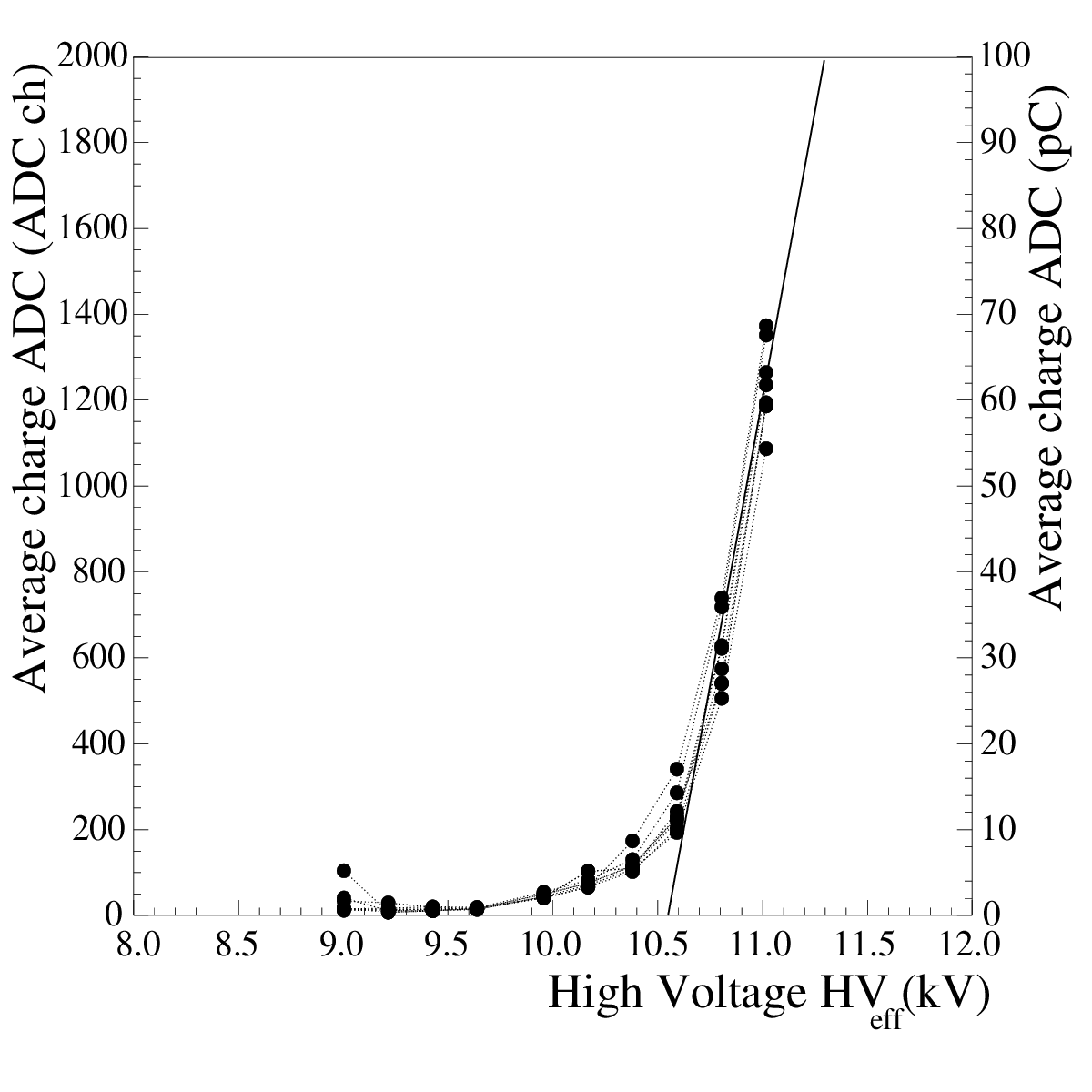}}
    \caption{Avarege avalanche charge of the eight monitor chamber signal as a 
function of HV$_{eff}$. 
     The slope is about 25 ADC ch/10 or 1.2pC/10V. Each point corresponds to 
10000 triggers.}
    \label{fig8}
  \end{figure}
\end{center}

In order to determine the sensitivity of GGM gaps to working point shifts, the 
avalanche to streamer transition was studied by two methods, the charge method 
and the efficiency method. In the charge method, the mean value of the ADC 
charge distribution in the whole ADC range is studied as a function of 
HV$_{eff}$ (fig.\ref{fig8}). Each point corresponds to 10000 events in the
whole ADC spectrum. In the plot three working point regions are identified
\begin{enumerate}
   \item inefficiency (HV$_{eff}<$ 9.7 kV);
   \item avalanche (9.7 kV $<$  HV$_{eff}<$ 10.6 kV;
   \item avalanche+streamer mode (HV$_{eff}>$ 10.6 kV).
\end{enumerate}
The best sensitivity to working point shifts is achieved in the 
avalanche+streamer region, estimated to be about 25~ADC~ch/10~V or 1.2pC/10V.

\begin{center}
  \begin{figure}[H]
    \resizebox{0.44\textwidth}{!}{\includegraphics{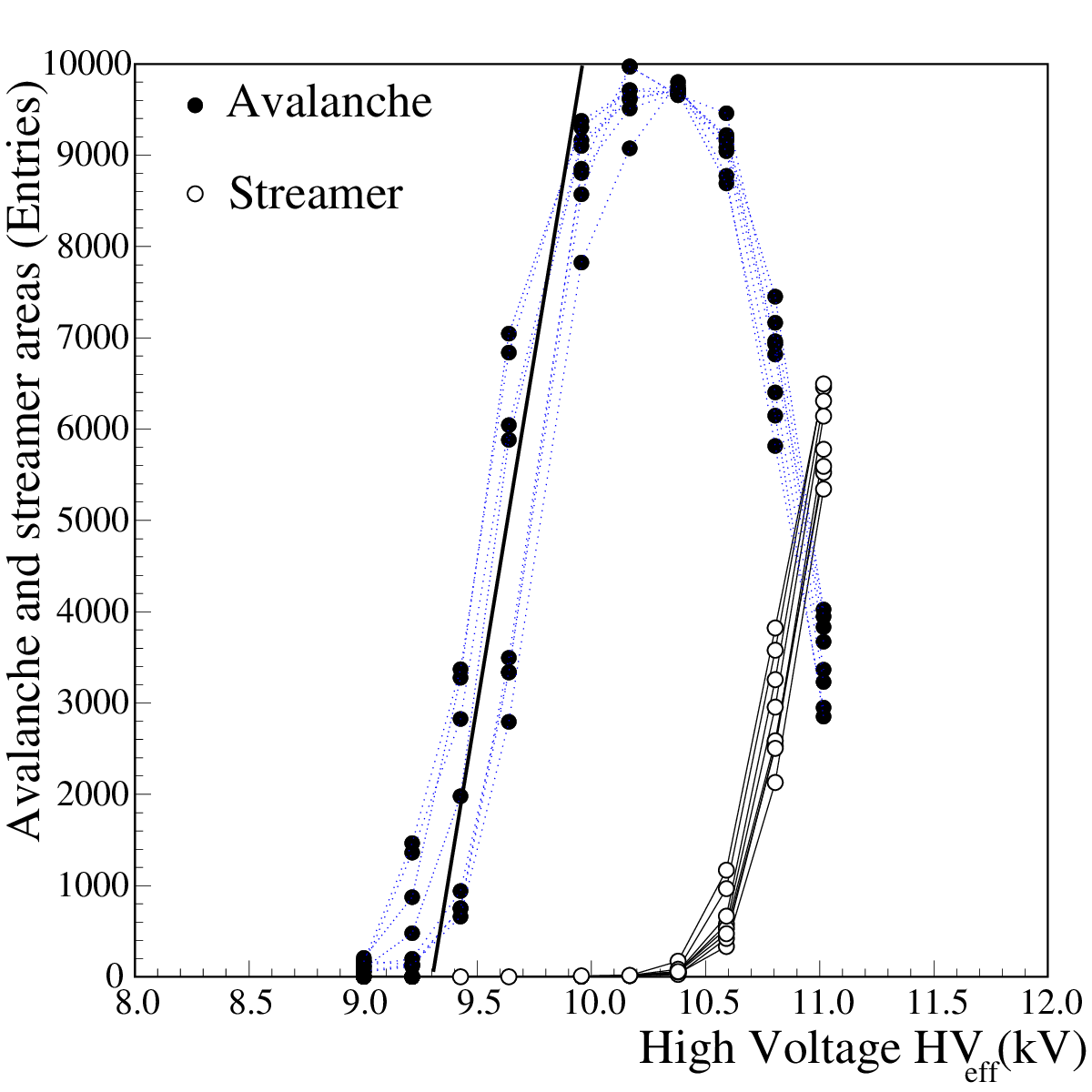}}
    \caption{Streamer and avalanche yields as a function of HV$_{eff}$. Each 
point corresponds to
      10000 collected triggers. The solid line has a slope of approximately 130 
events/10 V corresponding to a sensitivity of 1.3\%/10V.}
    \label{fig9}
  \end{figure}
\end{center}

In the efficiency method, the ADC avalanche event yield is studied
 as a function of HV$_{eff}$ (\ref{fig9}). The avalanche signal
 increases by increasing the HV applied to the gap, until it reaches a maximum 
value after which
the streamer component starts to increase. The 9.0kV-10.0kV shows a sensitivity 
to work point changes of approximately  1.3/\%/10V.

\section{Conclusions}
The  status of the  Gas Gain Monitoring System for the CMS RPC
Detector has been reported on. The purpose of GGM is to monitor any shift of the 
working point of the CMS
RPC detector. The GGM is being commissioned at CERN and is planned to start 
operation by the end of 2008. Preliminary results show good sensitivity to 
working point changes. Further tests are in progress to determine the 
sensitivity to gas variations.

\section{Acknowledgements}
 We warmly thank F.~Hahn and the CERN Gas Group for useful discussions and 
cooperation.

\end{document}